\documentclass[aps,prl,superscriptaddress,twocolumn]{revtex4}
\usepackage{amssymb}
\usepackage{graphicx}
\usepackage{amsmath,verbatim}
\usepackage[colorlinks=true,linkcolor=blue,citecolor=blue,urlcolor=blue]{hyperref}

\begin{document}

\title{Page Curve from Non-Markovianity}

\author{Kaixiang Su}
\affiliation{Institute for Advanced Study, Tsinghua University, Beijing 100084, China}
\affiliation{Department of Physics, University of California Santa Barbara, Santa Barbara, California, 93106, USA}

\author{Pengfei Zhang}
\email{pengfeizhang.physics@gmail.com}
\affiliation{Institute for Quantum Information and Matter, California Institute of Technology, Pasadena, California 91125, USA}
\affiliation{Walter Burke Institute for Theoretical Physics, California Institute of Technology, Pasadena, California 91125, USA}

\author{Hui Zhai}
\email{hzhai@tsinghua.edu.cn}
\affiliation{Institute for Advanced Study, Tsinghua University, Beijing 100084, China}
\date{\today}

\begin{abstract}

In this letter, we use the exactly solvable Sachdev-Ye-Kitaev model to address the issue of entropy dynamics when an interacting quantum system is coupled to a non-Markovian environment. We find that at the initial stage, the entropy always increases linearly matching the Markovian result. When the system thermalizes with the environment at a sufficiently long time, if the environment temperature is low and the coupling between system and environment is weak, then the total thermal entropy is low and the entanglement between system and environment is also weak, which yields a small system entropy in the long-time steady state. This manifestation of non-Markovian effects of the environment forces the entropy to decrease in the later stage, which yields the Page curve for the entropy dynamics. We argue that this physical scenario revealed by the exact solution of the Sachdev-Ye-Kitaev model is universally applicable for general chaotic quantum many-body systems and can be verified experimentally in near future.     

\end{abstract}

\maketitle

Studying open quantum many-body systems is of fundamental importance for understanding quantum matters and for future applications of quantum technology because all systems are inevitably in contact with environments, and decoherence due to coupling with environments is a major obstacle for future applications of quantum devices \cite{Preskill2018}. So far, most studies of open quantum systems are limited to either situation in which the systems are small or weakly correlated quantum many-body systems, or situations that the environment is treated by the Born-Markovian approximation \cite{scully1999quantum,breuer2002theory}. Little effort has been made on strongly correlated quantum many-body systems coupled to a non-Markovian environment. This is simply because both strong correlation and non-Markovianity are difficult to handle theoretically. 

\begin{figure}[t]
	\includegraphics[width=.95\columnwidth]{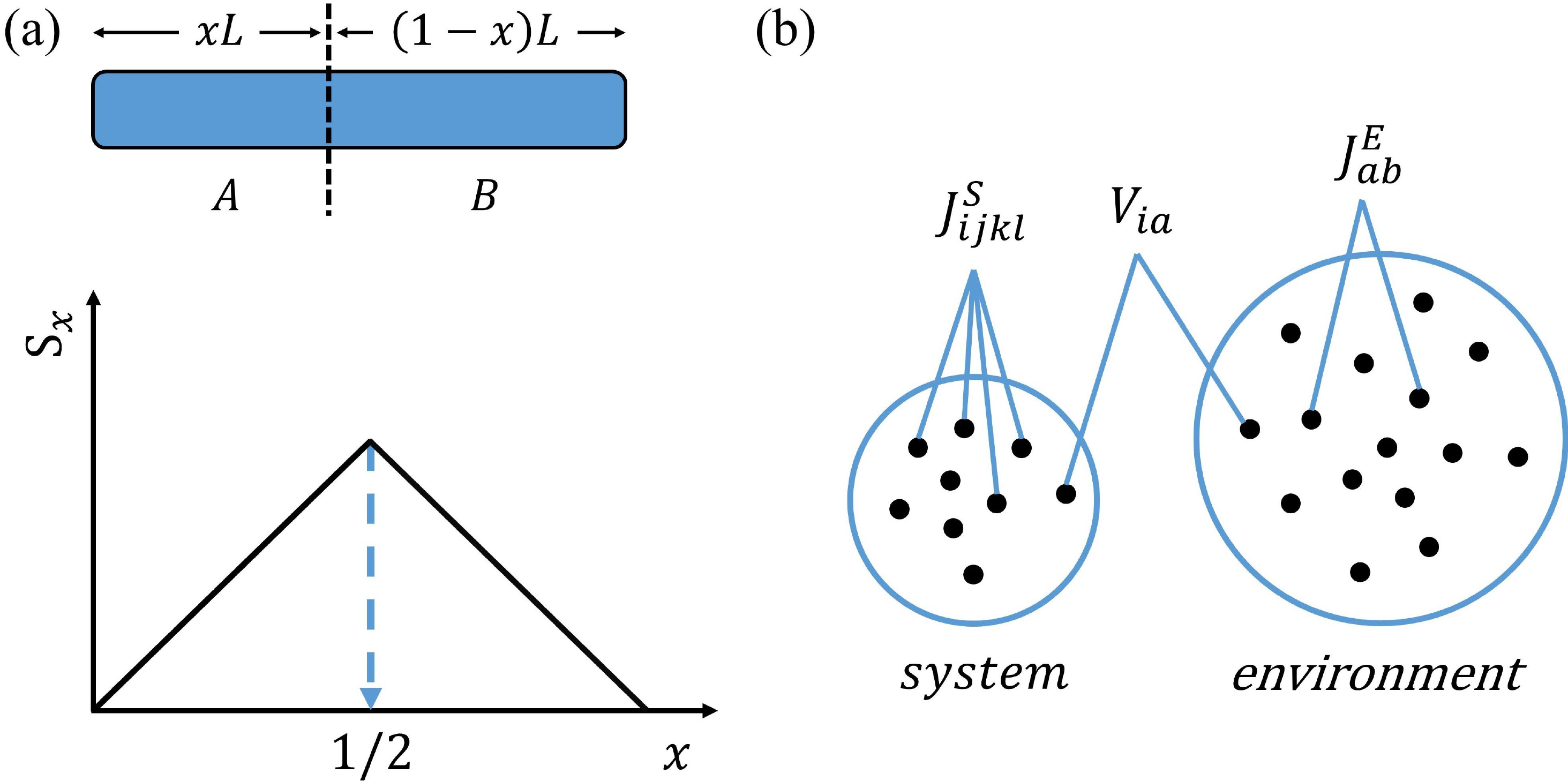}
	\caption{(a): Illustration of the Page curve: the entanglement entropy between two sub-systems with length $xL$ and $(1-x)L$. The total length is $L$ and $0\leqslant x \leqslant 1$. (b): Illustration of the setup: an SYK$_4$ system with $N$ Majorana fermions serves as the system and an SYK$_2$ system with $M$ Majorana fermions serves as the environment. Here $M\gg N$.} 
	\label{fig:illustration}
\end{figure}

Open systems are also of interest to gravity studies, and the best-known problem is the black hole information paradox \cite{hawking1975}. The central issue of the black hole information paradox is whether the black hole evaporation can be considered as undergoing unitary dynamics. If so, the entropy should first increase and then decrease as the black hole evaporates. Such an entropy curve as shown in Fig.~\ref{fig:illustration}(a) is known as the Page curve \cite{page1993}. Here the entanglement entropy is the entropy of the reduced density matrix of the radiation part $\mathcal{A}$ after tracing out the remaining black hole part $\mathcal{B}$ . As shown in Fig.~\ref{fig:illustration}(a), this entanglement entropy reaches the maximum when half of the black hole is evaporated, giving rise to the Page curve. Reproducing the Page curve from gravity theory is a challenging part of the black hole information problem, and progresses have been made recently using the semi-classical gravity calculations \cite{penington2019entanglement,almheiri2019entropy,almheiri2020page,almheiri2019replica,penington2019replica}.   

In this letter, we explore the Sachdev-Ye-Kitaev model \cite{maldacena2016remarks,Kitaev2018soft} with random four-Majorana fermions interactions (SYK$_4$), and this SYK$_4$ model is coupled to a system with random quadratic Majorana fermions couplings (SYK$_2$). The SYK$_2$ part contains a lot more degrees-of-freedom compared with the SYK$_4$ part such that the SYK$_2$ part can be viewed as the environment. The motivations for considering such a model are two folds. First, the SYK$_4$ model is exactly solvable in the large-N limit and its solution gives rise to a strongly correlated non-Fermi liquid state \cite{maldacena2016remarks}. Recently, techniques related to SYK$_4$ model has been widely used to construct exact solutions to address open issues of strongly correlated quantum matters \cite{gu2017local,davison2017thermoelectric,chen2017competition,song2017strongly,zhang2017dispersive,jian2017model,chen2017tunable,eberlein2017quantum,zhang2019evaporation,almheiri2019universal,gu2017spread,chen2020Replica,zhang2020,sk,Altman,sk jian,zhang2020entanglement,Liu2020non,Kitaev}. Here the situation we explored is also exactly solvable and we can use the solution to understand how a non-Markovian environment affects strongly correlated phases \cite{zhang2019evaporation,almheiri2019universal,chen2017tunable}. Secondly, the SYK$_4$ model is holographically dual to the Jackiw-Teitelboim gravity theory in the AdS$_2$ geometry with a black hole \cite{bulk spectrum Polchinski,bulk Yang,bulk2,bulk3,bulk4,bulk5}. Thus, the entropy dynamics of the SYK$_4$ system coupled to an environment \cite{gu2017spread,chen2020Replica,Kitaev} resembles the black hole evaporation process \cite{penington2019entanglement,almheiri2019entropy,almheiri2020page,almheiri2019replica,penington2019replica} and it will be interesting to study when a Page-like curve can emerge after turning on the coupling between the system and the environment. Remarkably, the main findings of this work bring these two aspects together, that is, we show that the Page curve emerges because of the non-Markovian effect of the environment.  

\textit{Model.} The system under consideration is illustrated in Fig.~\ref{fig:illustration}(b) and the total Hamiltonian is given by
 \begin{equation}\label{H1}
 \begin{aligned}
&\hat{H} = \hat{H}_\text{S}+\hat{H}_\text{E}+\hat{H}_{\text{SE}}\\
 &= \sum_{i<j<k<l}J^S_{ijkl}\psi_i \psi_j \psi_k \psi_l + i\sum_{a<b}J^\text{E}_{ab}\chi_a \chi_b + i\sum_{i,a}V_{ia}\psi_i \chi_a.
\end{aligned}
\end{equation}
Here $\psi_i$ ($i=1,\dots,N$) denotes $N$ Majorana fermions in the system and $\chi_a$ ($a=1,\dots,M$) denotes $M$ Majorana fermions in the environment, with $\{\psi_i,\psi_j\}=\delta_{ij}$ and  $\{\chi_a,\chi_b\}=\delta_{ab}$. Throughout the letter we will use the  subscript ``S" and ``E" to denote the system part and the environment part respectively. $\hat{H}_\text{S}$ and $\hat{H}_\text{E}$ are then SYK$_4$ and SYK$_2$ Hamiltonians. $J^S_{ijkl}$, $J^\text{E}_{ab}$ and $V_{ia}$ are independent random Gaussian variables with variances given by
\begin{equation}
\overline{(J^\text{S}_{ijkl})^2} = \frac{3!J_S^2}{N^3},\qquad \overline{(J^\text{E}_{ab})^2} = \frac{2!J_\text{E}^2}{M},\qquad \overline{(V_{ia})^2} = \frac{V^2}{M}.
\end{equation}
Throughout the paper, we take $J_S=1$ as the energy unit. We focus on the limit $M\gg N \gg 1$ in which the Schwinger-Dyson equation for $\chi$ contains no contribution from $\psi$, justifying that the $\chi$ part can be viewed as the environment. Therefore, the Green's function $G_\chi(\tau)=\left<T_\tau \chi_i(\tau)\chi_i(0)\right>$ of the environment takes the standard form of the SYK$_2$ model as \cite{maldacena2016remarks}
\begin{equation}\label{SYK2}
    G_\chi(\omega)=-\frac{2}{i\omega+i\text{sgn}(\omega)\sqrt{4J_\text{E}^2+\omega^2}}.
\end{equation}
The fact that this Green's function has frequency dependence means that the environment is treated beyond the Markovian approximation. 
 
We consider the situation in which the system and the environment is initially decoupled, and both are in thermal equilibrium with inverse temperatures $\beta_\text{S}$ and $\beta_\text{E}$ respectively. The initial density matrix is therefore given by $\rho(0)=
\frac{1}{Z_\text{S}Z_\text{E}}e^{-\beta_S \hat{H}_\text{S}}\otimes e^{-\beta_\text{E} \hat{H}_\text{E}}$ with $Z_\text{S}$ and $Z_\text{E}$ being the corresponding partition functions. Evolving the system with the Hamiltonian Eq.~\eqref{H1} and tracing out the environment, one obtains the reduced density matrix of the system $\rho_\text{S}(t)$ at time $t$ as 
\begin{equation}\label{entropy1}
\rho_S(t)=\text{tr}_\text{E}\left[e^{-i\hat{H}t}\rho(0)e^{i\hat{H}t}\right].
\end{equation}
The corresponding second R\'enyi entropy $S^{(2)}$ of the system is then given by
\begin{equation}\label{entropy1}
e^{-S^{(2)}_\text{S}(t)}= \text{tr}_\text{S}\left[ \rho_S(t)^2\right].
\end{equation}
Under the disorder replica diagonal assumption, $S^{(2)}_\text{S}(t)$ can be expressed in terms of path-integral over bilocal fields. In the large-$N$ limit, the integral is dominated by the saddle point solution, and the entropy can be obtained by evaluating the on-shell action.

\begin{figure}[t]
	\includegraphics[width=.95\columnwidth]{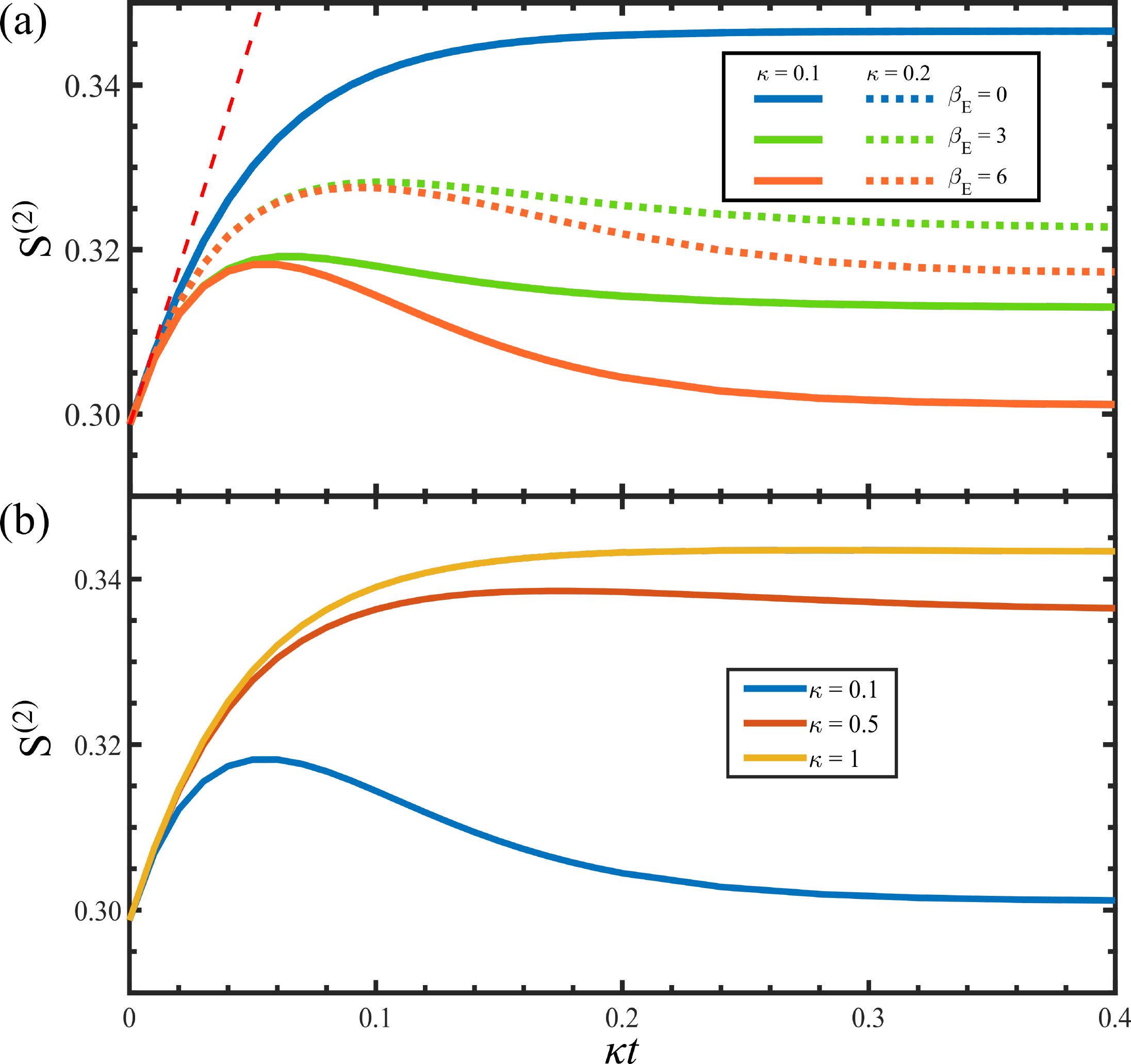}
	\caption{(a): Entropy curves for different $\kappa$ and $\beta_\text{E}$. Here $\kappa=0.1$ for solid lines and $\kappa = 0.2$ for dashed lines. Three different $\beta_\text{E}=(0, 3, 6)$ are plotted. The red dashed straight line represents the same initial slope for all curves. Two curves with $\beta_\text{E}=0$ in (a) coincide with the Markovian results. (b): Entropy curves with different $\kappa$ and a fixed $\beta_\text{E}=6$. The curve with large $\kappa$ in (b) coincides with $\beta_\text{E}=0$ curves in (a) and the Markovian results.} 
	\label{fig:entropy}
\end{figure}

\textit{Recovering the Markovian Results.} Below we will first discuss situations where the entropy dynamics of our model can recover the Markovian result. Here, by the Markovian result we mean dynamics obtained by solving the following Lindblad master equation \cite{scully1999quantum,breuer2002theory} 
\begin{equation}
\partial_t\hat{\rho} = -i[\hat{H}_\text{S},\hat{\rho}]+\sum_i\kappa_i\left(\hat{L}_i\rho \hat{L}_i^\dagger-\frac{1}{2}\{\hat{L}_i^\dagger \hat{L}_i,\hat{\rho}\}\right).\label{master}
\end{equation}
By treating the environment with the Markovian approximation, we only need to consider the Hamiltonian of the system and the dissipation operators, without having to explicitly include the environment. To make comparison with our model, we take $\kappa_i = {\kappa}$ and $\hat{L}_i = \psi_i$. Similar to previous procedures, we consider the initial thermal density matrix $\rho_\text{S}(0)=\frac{1}{Z_\text{S}}e^{-\beta_\text{S} \hat{H}_\text{S}}$, and we then evolve the system with Eq.~\eqref{master}. The second R\'enyi entropy can also be represented as a path-integral where the saddle points approximation is applicable. In the Markovian case, the R\'enyi entropy first grows linearly in time and then saturates to its maximum value $(N/2)\log 2$, forbidding the possibility of any page-like behaviors \cite{Zhai}. For reasons that will become clear below, we consider the large $J_\text{E}$ limit and fix $V^2/J_\text{E}$ as $\kappa$ in our model. Our discussions below will then identify the following conditions as sufficient for our model to recover the Markovian results. 

i) Infinite Environment Temperature. The SYK$_2$ environment becomes a Markovian one when $\beta_\text{E}=0$. This is because only the two-point function enters the effective action for entropy dynamics, and in the large $J_\text{E}$ limit, the Fourier transformation of the real-time Green's function of the environment $G_\text{E}^>(t,\beta)=\left<\chi_i(t)\chi_i(0)\right>_\beta/Z_B$ gives
\begin{equation}
    G^>_\text{E}(\omega,\beta_\text{E}) = \frac{1}{J_\text{E}}\frac{1}{1+e^{-\beta_E \omega}}.\label{Genvironment}
    \end{equation}
When $\beta_\text{E}=0$, the second term vanishes and the Green's function of the environment becomes frequency independent, which is equivalent to the Markovian approximation. Under this situation, the standard derivation of the master equation Eq. \ref{master} through a second-order perturbation theory yields the dissipation strength $\kappa=V^2/J_\text{E}$. As one can see in Fig. \ref{fig:entropy} (a), when $\beta_\text{E}$ is set to zero, the entropy curve becomes independent of $\kappa$. This universal curve also coincides with the results from the Markovian approximation.   

ii) Short Time. In the Markovian approach, it can be shown by perturbation that the entropy grows linearly at the initial stage, and when $\kappa t \ll 1$ \cite{npZhai,Zhai}, the growth rate can be derived analytically as 
\begin{equation}
  \frac{dS^{(2)}_\text{S}(t)}{dt}=\kappa N (1-2G_\text{S}^W(0,2\beta_S)). \label{shortMar1}
\end{equation}
Here we have defined the Wightman Green's function for the single SYK$_4$ model as 
\begin{equation}
G_\text{S}^W(t,\beta)=\frac{1}{Z_\text{S}}\left<\psi_i(t-i\frac{\beta}{2})\psi_i(0)\right>_\beta.
\end{equation} 
For our model, similar perturbative calculation \cite{Kitaev} for short-time yields
\begin{align}
    \frac{d S_\text{S}^{(2)}}{dt} = 
        2V^2N\int_{-t}^t dt'&\left[\left(G^>_\text{S}(t',2\beta_S) -G^W_\text{S}(t',2\beta_S) \right)\right.\nonumber\\
       &\left. \times G^>_\text{E}(t',\beta_B)\right].\label{short-nonMar}
\end{align}
By approximating $t^\prime=0$ in the integrand, Eq. \ref{short-nonMar} becomes
\begin{align}
    \frac{d S_\text{S}^{(2)}}{dt} = 
        \frac{V^2}{J_\text{E}}N (1-2G_\text{S}^W(0,2\beta_S)).\label{short-nonMar2}
\end{align}
By equalling $V^2/J_E=\kappa$, Eq. \ref{short-nonMar2} is the same as Eq. \ref{shortMar1}. This can also be seen in Fig. \ref{fig:entropy}(a) that the initial growth is linearly in $\kappa t$ and the slop is a constant for varying $\beta_\text{E}$ with fixed $\beta_\text{S}$. In other words, although the Green's function of the environment Eq. \ref{Genvironment} contains the frequency dependent part, it is not important for initial time and the short-time behavior is always dominated by the frequency independent part.  

iii) Large System-Environment Coupling. The entropy curve of our model also matches the Markovian result when $\kappa$ is sufficiently large compared with $J_\text{S}$. Since the short time limit is always Markovian as discussed in ii), here we focus on the long-time limit. Physically, when the coupling between system and environment is strong enough compared with the internal energy scales of the system, all Majorana fermions in the system tend to be maximally entangled with the environment, because the environment contains more degrees of freedom. Consequently, the entropy is expected to saturate to the maximum value $(N/2)\log 2$ in the long-time limit, which is the same as the Markovian case. This can also be shown more rigorously using the path-integral formalism by relating the R\'enyi entropy to the inner product of Kourkoulou-Maldacena pure states \cite{zhang2020entanglement,Liu2020non}. In Fig. \ref{fig:entropy}, one can see that the entropy curve gradually approaches the Markovian result as $\kappa$ increases.  

\begin{figure}[t]
    \includegraphics[width=.95\columnwidth]{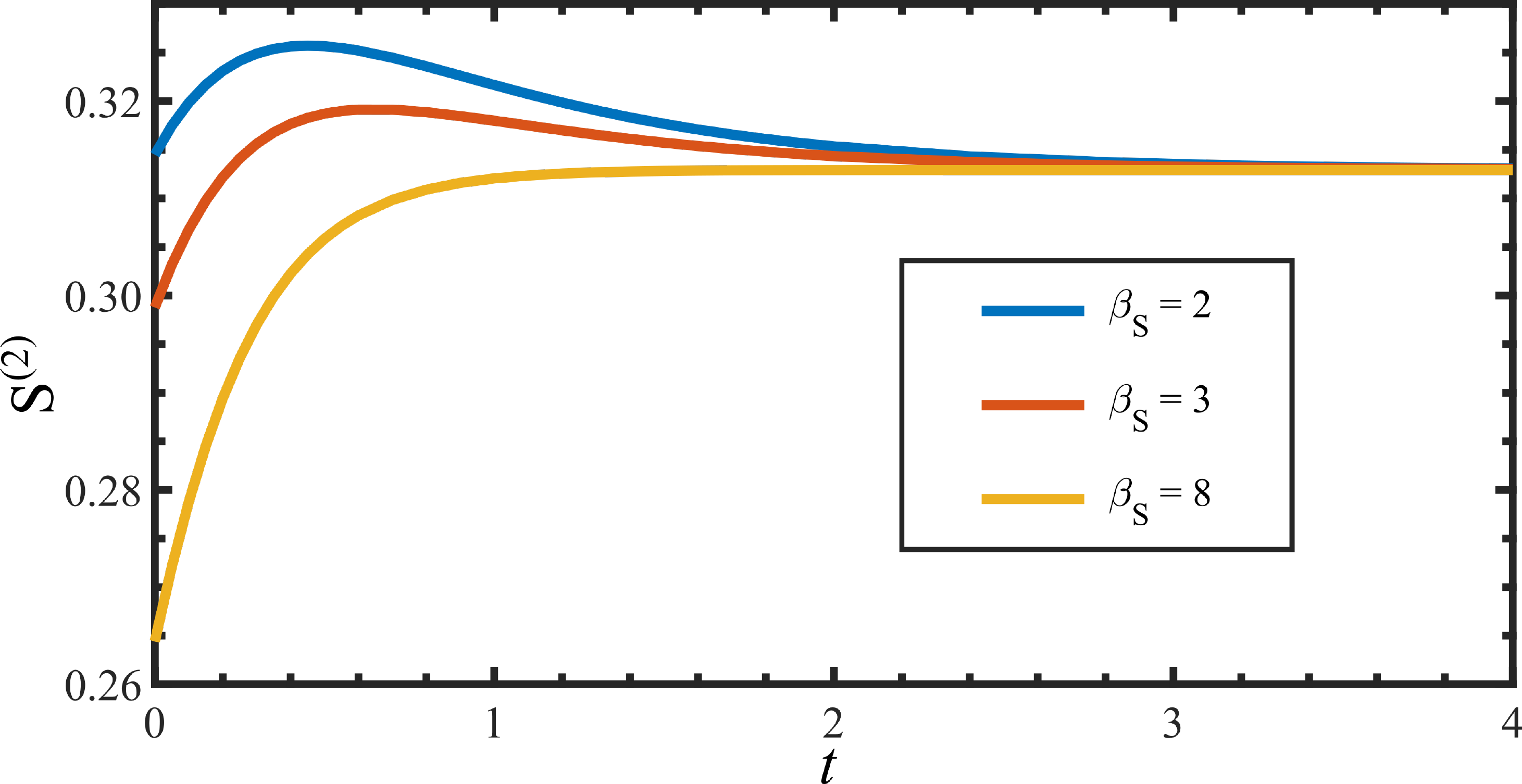}
    \caption{The entropy curve for varying $\beta_\text{S}$ with fixed $\kappa=0.1$ and $\beta_\text{E}=3$. Three different $\beta_\text{S}=(2,3,8)$ are plotted. Page-like behaviours is guaranteed when $\beta_\text{S}=2$ because the initial entropy is higher than the saturated entropy.} 
    \label{fig:page}
\end{figure}

\textit{The Page Curve.} Above we have shown that, under three situations, our model recovers the Markovian results, and the Markovian results do not display the Page curve for entropy dynamics. Hence, to reveal effects beyond the Markovian approximation, the following three conditions should be satisfied simultaneously, which are: i) the bath temperature should not be too high; ii) the evolution time should not be too short; iii) the coupling between system and environment should not be too large. Under these conditions, we find that Page curve for entropy dynamics is often observed, as was shown in Fig. \ref{fig:page}. Thus, this attributes the emergence of the Page curve to the \textit{beyond Markovian} effect. Since we have discussed that the entropy always increases at the initial time, it is essential to understand the decreasing behavior of the entropy curve at long time to understand the Page curve. Below we offer two physical understandings. 
 
The first understanding again relies on perturbation theory. When $\kappa$ is small, the entropy dynamics can be obtained by doing perturbation in $\kappa$, which yields the same perturbative results as Eq. \ref{short-nonMar}. Here, since we focus on the long-time behavior, we can simply replace $t$ by infinity and the range of integration in Eq. \ref{short-nonMar} is set to be $(-\infty,\infty)$. By expressing the Green's functions $G^>$ and $G^W$ in terms of the spectral functions $\rho$ \cite{maldacena2016remarks}
\begin{align}
&G^>(\omega,\beta)= \rho(\omega)\frac{1}{1+e^{-\beta \omega}} ,\\
&G^W(\omega,\beta)= \rho(\omega)\frac{1}{2\cosh{(\beta \omega/2)}},
\end{align}
and making use of the fact that $\rho(\omega)$ is even in $\omega$, we obtain the following expression: 
\begin{align}
  &  \frac{d S_\text{S}^{(2)}}{dt} = 2V^2N\int_{0}^\infty 
    \left[\frac{d\omega}{2\pi}\frac{\rho_\text{S}(\omega,2\beta_S)\rho_\text{E}(\omega,\beta_\text{E})}{2\cosh \beta_S \omega}\right.\nonumber\\
    &\left.\times \frac{(e^{\beta_S \omega}-1)(1-e^{(\beta_\text{E}-\beta_S)\omega})}{1+e^{\beta_\text{E} \omega}}\right]. \label{dSdtlong}
\end{align}
Note that in Eq. \ref{dSdtlong}, all terms are positive definite except for the $1-e^{(\beta_\text{E}-\beta_S)\omega}$ term. When the temperature of the environment is lower than the temperature of the system, $\beta_\text{E}>\beta_\text{S}$ and $1-e^{(\beta_\text{E}-\beta_S)\omega}<0$. Therefore, $d S_\text{S}^{(2)}/dt<0$ and the entropy decreases at long time, which yields the Page curve. This gives a sufficient condition for the emergence of the Page curve, that is, the small $\kappa$ and the lower environment temperature, which also agrees with the three aforementioned conditions. 

The second understanding replies on inspecting how the system entropy saturates at sufficiently long times. It is reasonable to assume that the system eventually reaches thermal equilibrium with the environment, and since the environment contains much more degrees-of-freedom, the saturation entropy is determined by $\kappa$ and $\beta_\text{E}$ and is independent of $\beta_\text{S}$. The saturation entropy is smaller when the environment temperature is lower, which corresponds to a decrease in thermal entropy. When the coupling $\kappa$ is smaller the saturation entropy is also smaller because this lowers the entanglement entropy.  On the other hand, the initial entropy of the system is mainly determined by parameters $J_\text{S}$ and $\beta_\text{S}$ of the system itself. When this saturation entropy is smaller than the initial entropy, the entropy has to decrease at a later stage, which also leads to a sufficient condition for the emergence of the Page curve. 

\textit{Summary.} In summary, we address the issue of when a Page curve can emerge in entropy dynamics of a system coupled to the environment. Although we use SYK model as an exactly solvable model to study this problem, the lesson we learn from our model reveals a general physical picture that should be applicable in generic chaotic quantum many-body systems. This physical picture contains two ingredients. First, at the initial stage, the entropy dynamics is always dominated by the Markovian process which leads to a linear increase of entropy in time. Secondly, a chaotic system thermalizes with the environment in the long-time limit. After thermalization, a low environment temperature and a weak system-environment coupling respectively suppress the thermal and the entanglement contributions to the system entropy, which ensures a lower system entropy at long time and forces the entropy to decrease at the later stage. The long-time decreasing behavior is essential for the emergence of the Page curve. This long-time behavior is distinct from the Markovian case where the system is often heated to infinite temperature and the long-time steady state is described by a density matrix given by the identity matrix. Therefore, the Page curve is a consequence of the non-Markovian environment. Since the entanglement entropy can now be measured experimentally \cite{Greiner1,Greiner2,Greiner3} and the coupling to the environment can be also highly controllable, for instance, in ultracold atomic systems, the physical picture revealed in this work can be experimentally verified in near future.

\textit{Acknowledgment.} This work is supported by Beijing Outstanding Young Scientist Program, NSFC Grant No. 11734010, MOST under Grant No. 2016YFA0301600.

\textit{Note added.} When finishing the manuscript, we become aware of a work by Chen in which the R\'enyi entropy dynamics has been stuided in a similar model by the perturbation theory \cite{Yu}.


\begin{thebibliography}{99}
\bibitem{Preskill2018}
J. Preskill, Quantum 2, 79 (2018).

\bibitem{scully1999quantum}
M. O. Scully, and M. S. Zubairy, Quantum Optics, Cambridge University Press, Cambridge, 1997.

\bibitem{breuer2002theory}
H. P. Breuer and F. Petruccione, The Theory of Open Quantum Systems, Oxford University Press, Oxford, 2007.

\bibitem{hawking1975}
S. W. Hawking, Commun.Math. Phys. \textbf{43} (1975) 199-220.                   

\bibitem{page1993}
D. N. Page, Phys. Rev. Lett. \textbf{71} (1993)1291.

\bibitem{penington2019entanglement}
G. Penington, J. High Energ. Phys. \textbf{2020}, 2 (2020).

\bibitem{almheiri2019entropy}
A. Almheiri, N. Engelhardt, D. Marolf, and H. Maxfield, J. High Energ. Phys. \textbf{2019} 63 (2019).

\bibitem{almheiri2020page}
A. Almheiri, R. Mahajan, J. Maldacena, and Y. Zhao, J. High Energ. Phys. \textbf{2020}, 149 (2020).

\bibitem{almheiri2019replica}
A. Almheiri, T. Hartman, J. Maldacena, E. Shaghoulian, and A. Tajdini, J. High Energ. Phys. \textbf{2020}, 13 (2020).

\bibitem{penington2019replica}
G. Penington, S. H. Shenker, D. Stanford, and Z. Yang, arXiv:1911.11977.

\bibitem{maldacena2016remarks}
J. Maldacena, and D. Stanford, Physical Review D \textbf{94} (2016) 106002.

\bibitem{Kitaev2018soft}
A. Kitaev, and S. Josephine Suh, J. High Energ. Phys. \textbf{2018}, 183 (2018).

\bibitem{gu2017local}
Y. {Gu}, X.-L. {Qi} and D. {Stanford}, 
  {{J. High Energ. Phys.} {\bfseries 2017} 125 (2017)}.

\bibitem{davison2017thermoelectric}
R. A. Davison, W. Fu, A. Georges, Y. Gu, K. Jensen and S. Sachdev, {{Phys. Rev. B} {\bfseries 95} 155131 (2017)}.

\bibitem{chen2017competition}
X. Chen, R. Fan, Y. Chen, H. Zhai and P. Zhang, {{Phys. Rev. Lett.} {\bfseries 119} 207603 (2017)}.
  
  \bibitem{Altman}
S. Banerjee and E. Altman, Phys. Rev. B \textbf{95}, 134302 (2017).
    
\bibitem{sk jian}
S.-K. Jian and H. Yao,  Phys. Rev. Lett. \textbf{119}, 206602 (2017).

\bibitem{song2017strongly}
X.-Y. Song, C.-M. Jian and L.~Balents, {Phys. Rev. Lett.} {\bfseries 119} 216601 (2017).

\bibitem{zhang2017dispersive}
P. Zhang, {Phys. Rev. B} {\bfseries 96} 205138 (2017).

\bibitem{jian2017model}
C.-M. Jian, Z. Bi and C. Xu, {Phys. Rev. B} {\bfseries 96} 115122 (2017).

\bibitem{eberlein2017quantum}
A. Eberlein, V. Kasper, S. Sachdev and J. Steinberg, {Phys. Rev. B}
  {\bfseries 96} 205123 (2017).

\bibitem{gu2017spread}
Y. {Gu}, A. {Lucas} and X.-L. {Qi}, J. High Energ. Phys. \textbf{2017}, 120 (2017).



\bibitem{chen2017tunable}
Y. {Chen}, H. {Zhai} and P. {Zhang}, {J. High Energ. Phys.} {\bfseries 2017} 150 (2017).

\bibitem{zhang2019evaporation}
P. Zhang, {Phys. Rev. B} {\bfseries 100} 245104 (2019).

\bibitem{almheiri2019universal}
A. {Almheiri}, A. {Milekhin} and B. {Swingle}, arXiv:1912.04912.

\bibitem{chen2020Replica}
Y. Chen, X.-L. Qi and P. Zhang, J. High Energ. Phys. \textbf{2020}, 121 (2020).

\bibitem{zhang2020entanglement}
P. Zhang, J. High Energ. Phys. \textbf{2020}, 2 (2020).

\bibitem{zhang2020}
P. Zhang, C. Liu, and X. Chen,  SciPost Phys. \textbf{8}, 094 (2020).

\bibitem{Liu2020non}
C. Liu, P. Zhang, and X. Chen, arXiv:2008.11955.

\bibitem{sk}
S.-K. Jian, and B. Swingle, arXiv:2011.08158.

\bibitem{Kitaev}
P. Dadras, A. Kitaev, arXiv: 2011.09622.


\bibitem{bulk Yang}
J. Maldacena, D. Stanford and Z. Yang, {Prog Theor Exp Phys} 2016 (12): 12C104.
    
\bibitem{bulk spectrum Polchinski}
J. Polchinski and V. Rosenhaus, J. High Energ. Phys. \textbf{2016}, 1 (2016).
    
\bibitem{bulk2}
K. Jensen, {Phys. Rev. Lett.} \textbf{117}, 111601 (2016).
    
\bibitem{bulk3}
A. Jevicki and K. Suzuki, J. High Energ. Phys. \textbf{2016}, 46 (2016).
    
\bibitem{bulk4}
G. Mandal, P. Nayak, and S. R. Wadia, {arXiv}:1702.04266.
    
\bibitem{bulk5}
D. J. Gross and V. Rosenhaus, J. High Energ. Phys. \textbf{2017}, 92 (2017).

\bibitem{Zhai}
Y. N. Zhou, L. Mao and H. Zhai, arXiv: 2101.*****.

\bibitem{npZhai}
L. Pan, X. Chen, Y. Chen, and H. Zhai, Nat. Phys. {\bf 16}, 767 (2020).

\bibitem{Greiner1}
R. Islam, R. Ma, P. M. Preiss, M. Eric Tai, A. Lukin, M. Rispoli, M. Greiner, Nature {\bf 528}, 77 (2015).

\bibitem{Greiner2}
A. M. Kaufman, M. Eric Tai, A. Lukin, M. Rispoli, R. Schittko, P. M. Preiss, M. Greiner, Science {\bf 353}, 794 (2016).

\bibitem{Greiner3}
A. Lukin, M. Rispoli, R. Schittko, M. Eric Tai, A. M. Kaufman, S. Choi, V. Khemani, J. L\'eonard, M. Greiner, Science {\bf 364}, 256 (2019).

\bibitem{Yu}
Y. Chen, arXiv: 2012.00223.


\end{thebibliography}
\end{document}